%Paper: hep-th/9507158
%From: Melanie Becker <mbecker@denali.physics.ucsb.edu>
%Date: Thu, 27 Jul 95 20:34:00 PDT
%Date (revised): Fri, 25 Aug 95 11:15:16 PDT
%Date (revised): Wed, 27 Sep 95 16:16:36 PDT

\input phyzzx
\Pubnum={\vbox{\hbox{NSF-ITP-95-62}\hbox{hep-th/9507158}}}
\date={}
\pubtype={}
\def\a{\alpha}
\def\b{\beta}
\def\g{\gamma}
\def\d{\delta}
\def\e{\epsilon}

\def\th{\theta}

\def\i{\iota}
\def\k{\kappa}
\def\l{\lambda}
\def\m{\mu}
\def\n{\nu}

\def\r{\rho}

\def\s{\sigma}

\def\c{\chi}

\def\G{\Gamma}
\def\D{\Delta}
\def\T{\Theta}

\def\S{\Sigma}

\def\O{\Omega}

\def\o{\over}
\def\ep{\epsilon}

\def\np{{\it Nucl. Phys. }}
\def\pl{{\it Phys. Lett. }}
\def\cmp{{\it Comm. Math. Phys. }}

\def\mpl{{\it Mod. Phys. Lett. }}
\def\prl{{\it Phys. Rev. Lett. }}
\def\pr{{\it Phys. Rev. }}

\def\no{\noindent}

\def\IR{\relax{\rm I\kern-.18em R}}
\font\cmss=cmss10 \font\cmsss=cmss10 at 7pt
\def\IZ{\relax\ifmmode\mathchoice
{\hbox{\cmss Z\kern-.4em Z}}{\hbox{\cmss Z\kern-.4em Z}}
{\lower.9pt\hbox{\cmsss Z\kern-.4em Z}}
{\lower1.2pt\hbox{\cmsss Z\kern-.4em Z}}\else{\cmss Z\kern-.4em Z}\fi}
\def\IN{\relax{\rm I\kern-.18em N}}
\def\IC{\relax{\rm I\kern-.18em C}}

\def\pa{\partial}
\def\tria{\triangleright}
\def\nsl{\nabla \!\!\!\! /}
\def\wid{\widehat}

\titlepage
\title{FIVEBRANES, MEMBRANES AND NON-PERTURBATIVE STRING THEORY}
\author{Katrin Becker}
\address{Institute for Theoretical Physics\break University of California
\break Santa Barbara, CA 93106-4030}
\author{Melanie Becker \break and \break Andrew Strominger}
\address{Department of Physics\break University of California \break
Santa Barbara, CA 93106-9530}

\abstract{Non-perturbative instanton corrections to the moduli space
geometry of type IIA string theory compactified on a Calabi-Yau space
are derived and found to contain order $e^{-1/g_s}$ contributions, where
$g_s$ is the string coupling. The computation reduces to a weighted sum
of supersymmetric extremal maps of strings, membranes and fivebranes into
the
Calabi-Yau space, all three of which enter on equal footing.
It is shown
that a supersymmetric 3-cycle is one for which the pullback of the K\"ahler
form vanishes and the pullback of the holomorphic three-form is a constant
multiple of the volume element. Quantum mirror
symmetry relates the sum in the IIA theory over supersymmetric,
odd-dimensional
cycles in the Calabi-Yau space to a sum in the IIB theory over
supersymmetric, even-dimensional cycles in the mirror. }

\endpage

\chapter{Introduction}
Progress in string theory hinges on gaining a better
understanding of non- perturbative effects. Recently there has been
substantial progress in understanding these effects in the context
of four-dimensional string theories with $N=2$ spacetime supersymmetry.
In this paper we will describe how non-perturbative corrections
modify the geometry of $D=4$, $N=2$ string theories. Our results also bear on
the role played by extended
objects other than strings in the theory which has -- perhaps
misleadingly -- come to be known as string theory.

The basic idea is as follows. IIA string theory contains membrane,
fourbrane and sixbrane solitons\footnote{1}{The standard, if confusing,
terminology is that a
$p$-brane is a $(p+1)$-dimensional extended object, a membrane is a twobrane
and
a string is a onebrane.} which carry Ramond-Ramond (RR) charges, while the IIB
theory contains string, threebrane and fivebrane solitons which carry
RR charges
\REF\hs{G.~T.~Horowitz and A.~Strominger, ``Black Strings and $p$-Branes'', \np
{\bf B360} (1991) 197.}
[\hs]. In the context of Calabi-Yau compactification,
euclidean $p$-branes can wrap around non-trivial $(p+1)$-cycles in the
Calabi-Yau space\footnote{2}{Wrapping modes of $p$-branes have been discussed
in a variety of related contexts in ref.
\REF\hetssl{A.~Strominger, ``Heterotic Solitons'', \np {\bf B343} (1990)
167. }
\REF\harliu{J.~A.~Harvey and J.~Liu, ``Magnetic Monopoles
in $N=4$ Supersymmetric Low-Energy Superstring Theory'',
\pl {\bf B268} (1991) 40.}
\REF\hutow{C.~M.~Hull and P.~K.~Townsend, ``Unity of Superstring Dualities'',
\np {\bf B438} (1995) 109, hep-th/9410167.}
\REF\town{P.~K.~Townsend, ``The Eleven-Dimensional Supermembrane Revisited'',
\pl {\bf B350} (1995) 184, hep-th/9501068.}
\REF\mbh{A.~Strominger, ``Massless Black Holes and Conifolds in
String Theory'', \np {\bf B451} (1995) 96, hep-th/9504090.}
\REF\tbh{B.~R.~Greene, D.~R.~Morrison and A.~Strominger,
``Black Hole Condensation and the Unification of String Vacua'',
\np {\bf B451} (1995) 109, hep-th/9504145.}
[\hetssl,\harliu,\hutow,\town,\mbh,\tbh].}. The supersymmetric, minimal action
configuration is an instanton, whose effects can be computed using standard
techniques. We shall find corrections which are independent of $g_s$,
the string coupling constant, and are just the standard worldsheet instanton
corrections. New corrections which behave as $e^{-1/g_s}$ are also found.
The existence of such effects was predicted in general from the growth of
string perturbation theory by Shenker
\REF\shenk{S.~H.~Shenker, ``The Strength of Non-Perturbative
Effects in String Theory'', presented at the Cargese Workshop
on Random Surfaces, Quantum
Gravity and Strings, Cargese, France, May 1990, preprint RU-90-47.} [\shenk]
, and in a context close to the present one by Witten
\REF\wit{E.~Witten, ``String Theory Dynamics in Various Dimensions'',
\np {\bf B443} (1995) 85, hep-th/9503124.} [\wit].

In $D=4$, $N=2$ supergravity, there are both hypermultiplets and
vector multiplets, but supersymmetry forbids them from talking
to one another at the level of the low-energy effective action
\REF\dlwp{B.~de Wit, P.~Lauwers and A.~Van Proeyen, ``Lagrangians
of $N=2$ Supergravity-Matter Systems'', \np {\bf B255} (1985) 569.}[\dlwp].
The dilaton ($e^{\phi}=g_s$) lives in a hypermultiplet. It follows that there
can be no non-perturbative corrections to the vector multiplet geometry.
However there can be, and we shall see that there are, non-perturbative
corrections to the geometry of the hypermultiplet moduli space.

In the IIA theory, hypermultiplets arise from the odd cohomology
(i.e. $H^3$) of the Calabi-Yau, while in the IIB theory they arise from
the even cohomology (i.e. $H^0$, $H^2$, $H^4$ and $H^6$).
In the IIA theory, we show that $e^{-1/g_s}$ corrections arise from membrane
instantons which wrap odd-dimensional cycles (i.e. 3-cycles) in
the Calabi-Yau, and therefore 	depend on the hypermultiplet
moduli which governs the size and shape of those cycles. In the IIB theory,
$e^{-1/g_s}$ corrections arise from string,
threebrane and fivebrane instantons which
wrap even (two-, four- and six-dimensional) cycles in the Calabi-Yau --
which for the IIB theory are the ones associated to hypermultiplets. The
existence of odd-(even-)dimensional $p$-branes in the
IIA (IIB) theory is in
perfect harmony with the  fact that only hypermultiplet geometry can be
corrected.
Thus the low-energy supergravity theory ``knows'' about
non-perturbative corrections.

In Yang-Mills theory, an instanton anti-instanton pair can be smoothly
deformed to the vacuum. Cluster decomposition can
then be used to determine the weighting of instanton sectors.
No such procedure is known for string theory or quantum
gravity in general, so some guesswork is involved in the rules stated
herein for inclusion of non-perturbative effects.
However they are severely constrained by consistency and other
considerations.

A first constraint is provided by mirror symmetry, which relates the
IIA theory compactified on a Calabi-Yau ${\cal X}$ to the IIB theory on
the mirror
${\widetilde {\cal X}}$ of ${\cal X}$. If this is to hold at the quantum
level, then the sum over odd cycles in the IIA theory on ${\cal X}$
must equal (in appropriate coordinates on the moduli spaces)
the sum over even cycles in the IIB theory on ${\widetilde {\cal X}}$.
The existence of such a relation seems mathematically surprising.

The second check, which is being investigated in
\REF\bbgms{K.~Becker, M.~Becker, B.~R.~Greene, D.~R.~Morrison
and A.~Strominger, in progress.}
[\bbgms], concerns conifold singularities
in the hypermultiplet moduli space and is discussed in section 6.

Ordinary mirror symmetry provides a check on the usual formulae
for worldsheet instanton corrections
\REF\dsww{X.~-G.~Wen and E.~Witten, ``World-Sheet Instantons and the
Peccei-Quinn Symmetry'', \pl {\bf B166} (1986) 397; M.~Dine, N.~Seiberg,
X.~-G.~Wen and E.~Witten, ``Non-Perturbative
Effects on the String World Sheet'', \np {\bf B278} (1986) 769;
``Nonperturbative Effects on the String World Sheet (II)'', \np {\bf B289}
(1987) 319.; J.~Distler and B.~R.~Greene, ``Some Exact Results on the
Superpotential From Calabi-Yau Compactification'', \np {\bf B309}
(1988) 295; P.~S.~Aspinwall and D.~R.~Morrison, ``Topological Field Theory
and Rational Curves'', \cmp {\bf 151} (1993) 245, hep-th/9110048.}
[\dsww] as was
spectacularly demonstrated in ref.
\REF\cgpd{P.~Candelas, X. de la Ossa, P.~S.~Green, L.~Parkes,
``A Pair of Calabi-Yau Manifolds as an Exactly
Solvable Superconformal Theory'', \np {\bf B359} (1991) 21.}
[\cgpd]. Second quantized mirror
symmetry provides an analogous check of the formulae herein. Proposed
examples
\REF\kv{S.~Kachru and C.~Vafa, ``Exact Results for $N=2$ Compactifications
of Heterotic Strings'', preprint HUTP-95-A016, hepth/9505105. }
\REF\fshv{S.~Ferrara, J.~A.~Harvey, A.~Strominger and C.~Vafa,
``Second Quantized Mirror Symmetry'', preprint EFI-95-26, hep-th/9505162.}
[\kv,\fshv],
relate $N=2$ heterotic and type II string compactifications. On
the heterotic side, the dilaton lies
in a vector multiplet and so hypermultiplet geometry can not
be renormalized. The tree-level hypermultiplet geometry is exact
and sums up type II RR instantons [\fshv]. A preliminary analysis
indicates that the appearance of $e^{-1/g_s}$ can be understood in this way
for dualities involving $K3$ fibrations
\REF\cvas{A.~Strominger and C.~Vafa, unpublished. }
[\cvas].

This paper is organized as follows. In preparation for the
analysis of the IIA theory, in section 2 we analyze membrane
instantons in Calabi-Yau compactification in the notationally simpler case
of $D=11$ supergravity. In 2.1 conditions for a supersymmetric map,
the analog of a holomorphic map for 3-cycles, is derived. In 2.2 we give an
explicit example of a supersymmetric map.
In 2.3 it is shown that such maps saturate
a Bogomol'nyi bound relating the membrane volume to the period
of the holomorphic three-form. In 2.4 we briefly review the quaternionic
geometry of the hypermultiplet moduli space. In 2.5 we construct
membrane
vertex operators (analogs of the usual string vertex operators),
count zero-modes and present formulae for the quantum corrections to the
geometry coming from supersymmetric 3-cycles. In 2.6 we argue that such
supersymmetric maps also count extremal black holes
in the IIB string theory.
In section 3 we obtain formulae in IIA string theory by compactifying
the eleven-dimensional theory to ten dimensions on a circle.
In this process a new type of membrane instanton, described in 3.2, arises in
which one leg of the membrane wraps the circle while the other two wrap
a Calabi-Yau 2-cycle. An important check on our results is that this reproduces
the standard formula for classical worldsheet instanton corrections.
The issue of multiple covers is raised but not resolved in 3.3. Fivebrane
instantons are discussed in section 4. In section 5
the IIB string is discussed, and it is argued that a sum over supersymmetric
maps
into even-dimensional Calabi-Yau cycles is related by mirror symmetry
to the
sum over odd-cycles in the IIA theory. The remaining steps to obtain
a precise mathematical form of this relation are outlined. In section 6
we briefly address the problem of conifold singularities in the
hypermultiplet moduli space.
We conclude
with a discussion on the role of $p$-branes
in string theory in section 7. Our conventions are in an appendix.

The derivation of the corrections from supersymmetric maps of
membranes into 3-cycles in sections 2 and 3 was greatly aided by the
prior construction
\REF\smb{E.~Bergshoeff, E.~Sezgin and P.~W.~Townsend,
``Supermembranes and Eleven-Dimensional
Supergravity'', \pl
{\bf B189} (1987) 75; E.~Bergshoeff, E.~Sezgin, P.~K.~Townsend, ``Properties
of the Eleven-Dimensional Super-Membrane Theory'', {\it Ann. Phys.} {\bf
185} (1988) 330. }
[\smb] of the covariant supermembrane
action. Such actions have not been constructed for the $p$-branes
of the IIB theory. Our formulae in section 4 are accordingly
somewhat preliminary and schematic. Indeed many aspects of
instanton corrections for both IIA and IIB theories
-- such as multiple covers, topological sums and
zero modes -- are not settled herein. Further details and analysis will be
presented elsewhere.

\chapter{Non-Perturbative Effects in D=11 Supergravity}

\no In this section we consider Calabi-Yau compactification to five dimensions
of eleven-dimensional supergravity. We will see that the geometry of this
five-dimensional theory is non-perturbatively corrected by instantons,
corresponding to supermembranes wrapping around 3-cycles
of the Calabi-Yau space.

\section{Membrane Instantons}
\no The bosonic part of the euclidean action of $D=11$ supergravity
\REF\nahm{W.~Nahm, ``Supersymmetries and their Representations'',
\np {\bf B135} (1978) 149.}
\REF\cjs{E.~Cremmer, B.~Julia and J.~Scherk, ``Supergravity Theory
in Eleven Dimensions'', \pl {\bf B76} (1978) 409.}
[\nahm,\cjs] is
$$
{\cal S}_{11}=
{1\o 2\pi^2 \ell^9 } \int d^{11} x \sqrt{{\wid g}} \left[- {\wid R}+{1\o 48  }
( dC) ^2\right]+  {i \o 12 \pi ^2 \ell^2} \int C\wedge dC\wedge dC.
\eqn\eleac
$$
In this expression, ${\wid g}$
is the spacetime metric (the hat denotes eleven-dimensional quantities),
$C$ is a three-form potential and $\ell$ is the eleven-dimensional
Planck length.
The simplest way to analyze the membrane
instanton solutions of this theory is to go directly to an
effective description on scales large relative to the thickness of the
membrane, in which the membrane is described by an effective
worldbrane action. The form of this action is completely fixed by
supersymmetry as
[\smb] \footnote{3}{The construction of the membrane action
built on earlier work
of Hughes, Liu and Polchinski \REF\hlp{J.~Hughes, J.~Liu and J.~Polchinski,
``Supermembranes'', \pl {\bf
B180} (1986) 370. }[\hlp]. The membrane soliton solution was derived in ref.
\REF\duste{M.~J.~Duff and K.~S.~Stelle, ``Multimembrane
Solutions of $D=11$ Supergravity'', \pl {\bf B253} (1991) 113.}
[\duste] and the quantization condition for the membrane tension
is discussed in ref.
\REF\dulimi{M.~J.~Duff, J.~T.~Liu and R.~Minasian, ``Eleven-Dimensional Origin
of String/String Duality: A One Loop Test'',
preprint CTP-TAMU-26/95, hep-th/9506126.}
[\dulimi]. A good review article is ref.
\REF\tls{P.~K.~Townsend, ``Three Lectures on Supermembranes'',
in Proceedings of the ``Trieste Spring School'', 11-19 April 1988.}
[\tls]. In the euclidean version
presented here, there is a doubling of the fermionic components
(because the Majorana condition cannot be imposed in eleven-dimensional
euclidean space),
so only half of the components of $\T$ should be integrated over, as
in the four-dimensional discussion in ref.
\REF\hlf{I.~Affleck, J.~Harvey and E.~Witten ,``Instantons and (Super)Symmetry
Breaking'', \np {\bf B206} (1982) 413; E.~Witten, ``An $SU(2)$ Anomaly'',
\pl {\bf B117} (1982) 324.}
[\hlf].}:
$$
\eqalign{
{\cal S}_3={1\o \ell^3 } \int
 d^3 \s & \sqrt{h}\Bigl[ {1\o 2} h^{\a\b} \pa_{\a} X^M
\pa_{\b}X^N {\wid g}_{MN}-{1\o 2} -i
{\bar \T }  \G^{\a} \nabla _{\a} \T \cr
& +{i\o 3!}\e^{\a\b\g} C_{MNP} \pa_{\a} X^M
\pa_{\b}X^N \pa_{\g}X^P + \dots \Bigr] .\cr}
\eqn\sma
$$
Here $X^M(\s)$, with $M,N=1,\dots,11$,
describes the membrane configuration, $\T$ is an
eleven-dimensional Dirac spinor and $h_{\a\b}$ with $\a,\b=1,2,3$,
is an auxiliary worldbrane metric with euclidean signature. In this, and
all subsequent expressions only the leading order terms in powers of the fermi
fields are displayed. The above action is a generalization of the Green-Schwarz
action for superstrings  \REF\gs{M.~B.~Green
and J.~H.~Schwarz, ``Properties of the Covariant Formulation of Superstring
Theories'',
\np {\bf B243} (1984) 285; M.~B.~Green and J.~H.~Schwarz, ``Covariant
Description of
Superstrings'', \pl {\bf B136} (1984) 367.}
[\gs], to a two dimensional extended object. We henceforth adopt units
in which $\ell=1$.

The equation of motion of the auxiliary worldbrane metric sets it equal to the
induced metric
$$
h_{\a\b}=\pa_{\a}X^M \pa_{\b} X^N {\wid g}_{MN}.
\eqn\ei
$$
In the following, we will consider field configurations where $dC=0$,
which is required for supersymmetric compactifications
\REF\candelas{P.~Candelas and D.~J.~Raine, ``Compactification
and Supersymmetry in $D=11$ Supergravity'',
\np {\bf B248} (1984) 415.}[\candelas].
The first term of the action then reduces to the volume of the membrane,
and the $X$ equation of motion requires extremization of the volume.

The global fermionic symmetries act on
the membrane fields as
$$
\eqalign{
&\d_{\ep}\T=\ep, \cr
&\d_{\e} X^M=i { \bar \e} \G^M \T, \cr}
\eqn\strn
$$
where $\e$ is a constant anticommuting eleven-dimensional spinor.
The theory is invariant under local fermionic transformations,
the so-called $\k$ symmetries, which act on the fields as
$$
\eqalign{
&\d_{\k} \T=2P_+ \k(\s) , \cr
&\d_{\k} X^M =2i{\bar \T} \G^M P_+ \k(\s),\cr}
\eqn\ktrn
$$
where $\k$ is a $D=11$ spinor and $P_{\pm}$ are the projection operators
[\hlp]\footnote{4}{The factor of $i$, which does not appear in ref. [\tls],
arises in euclidean space.}:
$$
P_{\pm}={1\o 2} \left( 1\pm { i\o 3! } \e^{\a\b\g} \pa_{\a} X^M\pa_{\b} X^N
\pa_{\g} X^P \G_{MNP}\right).
\eqn\pdef
$$
These obey
$$ \eqalign{
&P_{\pm}^2=P_{\pm}, \cr
&P_+ P_-=0,\cr
&P_++P_-=1.\cr}
\eqn\pprp
$$

A general bosonic membrane configuration $X(\s)$ breaks all the global
supersymmetries generated by $\ep$. Unbroken supersymmetries remain
if and only if {\strn} can be compensated
for by a $\k$-transformation, i.e. there exists a spinor $\k(\s)$ such that
$$
\d_{\e} \T+\d_{\k} \T= \e+2P_+ \k(\s)=0.
\eqn\keps
$$
Applying $P_-$ to both sides one finds that this requires
$$
P_-\ep=0.
\eqn\peps
$$

We are interested in the case for which six dimensions are
compactified on a Calabi-Yau manifold, and the remaining are flat (this has
been
studied in ref.
\REF\duffpro{M.~J.~Duff, ``Kaluza-Klein Theories and
Superstrings'', in
Architecture of the Fundamental
Interactions at Short Distances, Proceedings of the
{\it ``Les Houches Summer School''}, July 1985, Eds. P.~Ramond
and R.~Stora, North-Holland, 1987.}
\REF\hoto{P.~S.~Howe and P.~K.~Townsend, ``Supermembranes and the Modulus
Space of $d=4$ Superstrings'', in {\it Supermembranes and Physics in $2+1$
dimensions}, eds. M.~J.~Duff, C.~N.~Pope and E.~Sezgin (World Scientific
1990).}
\REF\cad{A.~Cadavid, A.~Ceresole, R.~D'~Auria and S.~Ferrara, ``
11-Dimensional Supergravity Compactified on Calabi-Yau Threefolds'',
preprint CERN-TH/95-166, hep-th/9506144.}
\REF\pato{G.~Papadopoulos and P.~K.~Townsend, ``Compactification
of $D=11$ Supergravity on Spaces of Exceptional Holonomy'', preprint
R/95/31, hep-th/9506150.}
[\duffpro,\hoto,\cad,\pato] among others). In complex coordinates the metric is
of the mixed form
${\wid g}_{m{\bar n}}$ with $m,n=1,2,3$. A Calabi-Yau manifold
admits a nowhere vanishing holomorphic  $(3,0)$ form
$$
\O={1\o 3!} \O_{mnp}(X)dX^m\wedge dX^n \wedge dX^p,
\eqn\eai
$$
and a K\"ahler form ${\wid J}_{m{\bar n}}=i {\wid g}_{m{\bar n}}$. There are
eight
covariantly constant spinors $\e$,
associated to the global supersymmetries which remain unbroken by
the Calabi-Yau compactification.

To analyze the condition {\peps} further, let $\e_+=(\e_-)^*$ (in an
imaginary representation of the gamma matrices) be two
covariantly constant six-dimensional spinors with opposite chirality,
where $\e_+$ is defined to have positive chirality.
The normalization can be chosen so that:
$$
\eqalign{
&\g_{mnp}\e_+=e^{-{\cal K}}\O_{mnp}\e_-,\cr
&\g_{{\bar m} np} \e_+=2i{\widehat J}_{{\bar m}[n}\g_{p]} \e_+,\cr
&\g_{{\bar m}} \e_+=0,\cr}
\eqn\ercl
$$
where we have introduced
$$
{\cal K}={1\o 2}\left({\cal K}_{\cal V}-{\cal K}_{\cal H}\right),
\eqn\kkk
$$
with the K\"ahler potential on the moduli space
of complex structures given by
$$
{\cal K}_{\cal H}=-\log\left( i \int\nolimits_{\cal X}
\O \wedge {\bar \O}\right) ,
\eqn\kh
$$
and the potential on the K\"ahler moduli space
$$
{\cal K}_{\cal V}=-\log \left( {4\o 3 } \int\nolimits_{\cal X}  {\wid J} \wedge
{\wid J} \wedge {\wid J}\right) .
\eqn\kv
$$
Define
$$
\ep_{\theta}=e^{i \th}\e_+ +e^{-i\th} \e_-,
\eqn\eii
$$
and let $\e$ in eq. {\peps} be of the form $\e_{\theta} \l$, with $\l$ a
five-dimensional spinor. One then finds that $P_- \e_{\theta}=0$ implies
$$
\eqalign{&{i \o 3! } \e^{\a\b\g} \pa_{\a} X^m\pa_{\b} X^n
\pa_{\g} X^p e^{i\theta} e^{-{\cal K}}
\O_{mnp} \e_-
-e^{-i\th} \e_-\cr
&+\e^{\a\b\g} \pa_{\a}X^m \pa_{\b} X^{{\bar n}}{\wid J}_{m{\bar n}}
\pa_{\g} X^{\bar p} \g_{\bar p} e^{-i\th} \e_-+{\rm c.c.}=0.\cr}
\eqn\gps
$$
The spinors $\e_-$ and $\g_{\bar p} \e_-$ are linearly independent since
they transform differently under the holonomy group. Therefore,
the third term cannot be cancelled by anything else and must vanish
on its own. This requires that the pullback of the K\"ahler form on
to the membrane vanishes:
$$
\pa_{[\a}X^m \pa_{\b]}X^{{\bar n}}{\wid J}_{m{\bar n}}=0.
\eqn\jpl
$$
The remaining terms can cancel if and only if the phase and magnitude
of the first are constant. This cancellation requires:
$$
\pa_{\a} X^m \pa_{\b} X^n \pa_{\g} X^p \O_{mnp}=e^{-i\varphi}
e^{\cal K}\e_{\a\b\g},
\eqn\opl
$$
where $\varphi=2\theta+\pi/2$. This equation states that the membrane volume
element is
proportional to the pullback of $\O$.
Given {\jpl} and {\opl}, a spinor
of the form $P_+\e$ is covariantly constant and obeys {\peps}. Hence {\jpl} and
{\opl}
are the necessary and
sufficient conditions for a map $X(\s)$ to be supersymmetric
\footnote{5}{We are grateful to E.~Witten and S.~-T.~Yau for very helpful
discussions
on these conditions.}.

\section{A Supersymmetric 3-Cycle in the Quintic}
\no It is easy to see that a flat membrane in $\IR^{10}$ is supersymmetric.
A more nontrivial example of a solution of eqs. {\jpl} and {\opl}, suggested
to us by G.~Moore
\REF\grmo{G.~Moore, private communication.}
[\grmo] and by E.~Witten
\REF\wittwo{E.~Witten, private communication.}[\wittwo], is as
follows.

The equation
$$
\sum_{i=1}^5 ({X^i})^5=0,
\eqn\qui
$$
defines a quintic hypersurface in ${\rm C P}^4$. The Ricci-flat
metric ${\wid g}$ on the quintic has the isometry
$$
X^i{\buildrel D \over \longmapsto} X^{\bar \i}.
\eqn\quii
$$
$D$ leaves fixed a three-dimensional submanifold
$ C_3$ on which all the ${X^i}$'s are real. We now show that $C_3$ is a
supersymmetric 3-cycle.

Since $D$ preserves the metric but reverses the orientation, it
acts on the K\"ahler form associated to ${\wid g}$ as
$$
{\wid J} {\buildrel D \over \longmapsto}-{\wid J}.
\eqn\quiii
$$
On the other hand the pullback of ${\wid J}$ onto the fixed surface
$C_3$ must be invariant. This is only possible if the pullback vanishes.
It follows that {\jpl} is satisfied.

To see that the pullback of $\O$ is the induced volume form on $C_3$,
we note that the holomorphic 3-form can be written as
\REF\aswit{A.~Strominger and E.~Witten,
``New Manifolds for Superstring Compactification'', \cmp {\bf 101}
(1985) 341.}
$$
\O={dY^1\wedge dY^2\wedge dY^3 \o (Y^4)^4},
\eqn\quiv
$$
where we have introduced the inhomogeneous coordinates:
$$
Y^k={X^k \o X^5} \qquad {\rm with } \qquad k=1,\dots,4.
\eqn\yxx
$$
The norm of $\O$ is
$$
\parallel \O \parallel ^2={1\o 3!} \O_{mnp}{\bar \O}^{mnp}=
{1 \o {\wid g} |Y^4|^8   },
\eqn\quv
$$
where ${\wid g}$ is the determinant of $\wid g_{m{\bar n}}$.
Since $\O$ is covariantly constant, this norm is a constant
$$
\parallel \O \parallel ^2=8e^{2{\cal K}}.
\eqn\quvi
$$
It follows that
$$
{\wid g}= {e^{-2{\cal K}}\o 8|Y^4|^8}.
\eqn\quvii
$$
When the pullback of ${\wid J}$ vanishes, the induced metric is
$$
h_{\a\b}=2\pa_{\a} Y^m \pa_{\b} Y^{\bar n} {\wid g}_{m{\bar n}}
\eqn\quviii
$$
and, using {\quvii} and reality of $Y^4$ on $C_3$
$$
\sqrt{h}=|\det(\pa Y)|  {e^{-{\cal K}}\o |Y^4|^4} .
\eqn\quix
$$
Substituting into eq. {\quiv} and pulling back to $C_3$ one finds
$$
{1\o 3!} d\s^{\a} \wedge d\s^{\b} \wedge d\s^{\g}
\pa_{\a} Y^m \pa_{\b} Y^n \pa_{\g} Y^p \O_{mnp}=
e^{i \omega} e^{\cal K}\sqrt{h} d\s^1\wedge d\s^2 \wedge d\s^3,
\eqn\qux
$$
for some phase $\omega$. Therefore eq. {\opl} is satisfied.

\section{A Bogomol'nyi Bound}

\no Eqs. {\jpl} and {\opl} imply that the membrane
has minimized its volume. To see this consider
$$
\int\nolimits d^3 \s \sqrt{h}
{\bar \e}_{\theta} P_-^{\dagger}  P_-\e_{\theta}
\geq 0,
\eqn\dthree
$$
with $P_-$ constructed from six-dimensional gamma matrices.
Using the relation $P_- ^{\dagger} P_-=P_-$ the inequality becomes
$$
2{\wid V_3}\geq
e^{-{\cal K}} \left( e^{i\varphi}\int
 \O +e^{-i\varphi} \int {\bar \O}  \right) ,
\eqn\dint
$$
where $\varphi$ is defined below eq. {\opl}.
By adjusting $\varphi$ to maximize the right hand side one finds
\footnote{6}{The factor of $e^{{\cal K}_{\cal V}/2} $
is absent in a related bound discussed
in ref.
\REF\cafv{A.~Ceresole, R.~D'Auria, S.~Ferrara and A.~Van Proeyen,
``Duality Transformations in Supersymmetric Yang-Mills Theory Coupled
to Supergravity'', \np {\bf B444} (1995) 92, hep-th/9502072.}
[\cafv,\mbh] because of rescaling of the four-dimensional metric. }
$$
{\wid V_3}\geq e^{-{\cal K}} \mid \int \O \mid.
\eqn\ubnd
$$
On the other hand the left hand side of {\dthree} vanishes if and only
if $P_-\e_{\theta}=0$, so this bound is saturated if and only if
the maps obey the supersymmetry conditions {\jpl} and {\opl}.

\section{Quaternionic Geometry and the Five-Dimensional Action}
\no We wish to compute the instanton corrections to the low-energy effective
action in five dimensions. In this subsection we recall the relevant features
of this action. It contains both hypermultiplets and vector multiplets. $N=2$
supersymmetry forbids neutral couplings (for the lowest dimension terms)
between
these multiplets. The moduli space ${\cal M}$ is a direct product of the moduli
space of hypermultiplets ${\cal M_H}$ and the moduli space of vector multiplets
${\cal M_V}$:
$$
{\cal M}={\cal M_H}\times {\cal M_V}.
\eqn\prod
$$
Since the instanton action scales with the size of the Calabi-Yau space,
which is parametrized by a hypermultiplet, instanton corrections can affect
only
the hypermultiplet geometry.

As shown in ref.
\REF\bagwit{J.~Bagger and E.~Witten, ``Matter Couplings in
$N=2$ Supergravity'', \np {\bf B222} (1983) 1.}
[\bagwit] in the context of $N=2$, $D=4$ supergravity, the $4n$ scalars of $n$
hypermultiplets parametrize a quaternionic geometry, with
holonomy group $Sp(n)\cdot Sp(1)$ and a non-vanishing $Sp(1)$
connection\footnote{7}{An explicit construction of the quaternionic
manifolds that arise at tree level
in type II string theories compactified on Calabi-Yau manifolds
has been carried out
in ref. \REF\fesa{S.~Ferrara and S.~Sabharwal, ``Quaternionic Manifolds
for Type II Superstring Vacua on Calabi-Yau Spaces'', \np {\bf B332}
(1990) 317.}[\fesa].}. The Riemann tensor of this geometry can be obtained
from:
$$
R_{ijkl}\g_{C I}^l \g _{BJ}^k=\e_{CB} {\cal R}_{ijIJ} +\e_{IJ}
{\cal R}_{ij CB},
\eqn\rite
$$
where $C,B=1,2$, $I,J=1,\dots, 2n$ and $i,j=1,\dots,4n $;
${\cal R}_{ijAB}$ and
${\cal R}_{ijIJ}$ are the $Sp(1)$ and $Sp(n)$
curvatures respectively and $\g_{AJ}^i$ are covariantly
constant functions of the $4n$ scalars that satisfy identities
similar to those of the Dirac gamma matrices.
The $Sp(1)$ connection can be written in the form
$$
{\cal R}_{ijAB}=\varsigma
^2\left(\g_{iAI}{\g_{jB}}^I-\g_{jAI}{\g_{iB}}^I \right)
\eqn\rgggg
$$
while the $Sp(n)$ connection is given by
$$
{\cal R}_{ijIJ}=\varsigma ^2\left( \g_{iAI} {\g_{j}^A}_J-
\g_{jAI}{\g_{i}^A}_J\right) +\g_i^{AL} {\g_{jA}}^K {\cal R}_{IJKL},
\eqn\spnr
$$
where ${\cal R}_{IJKL}$ is totally symmetric in its indices and $\varsigma^2$
is proportional to the five-dimensional Newton's constant.

In the following we shall focus on corrections to the
four-fermi coupling, given by
 \REF\gst{
M.~G\"unaydin, G.~Sierra and P.~K.~Townsend, ``The Geometry of
$N=2$ Maxwell-Einstein Supergravity and Jordan Algebras'', \np {\bf B242}
(1984)
244; ``Gauging the $D=5$ Maxwell-Einstein Supergravity
Theories: More on Jordan Algebras'', \np {\bf B253} (1985) 573;
G.~Sierra, ``$N=2$ Maxwell Matter Einstein Supergravities in
$D=5$, $D=4$ and $D=3$'', \pl {\bf B157} (1985) 379.}[\gst]
$$
\int d^5 x \sqrt{{\wid g}}( {\bar \c}^I \c^J)
( {\bar \c}^K \c^L ){\cal R}_{IJKL} ,
\eqn\fffm
$$
where $\c^I$ is the fermionic component of the hypermultiplet. We note
that, since the spinor $\c^I$ has four real (anticommuting)
components, and ${\cal R}_{IJKL}$ is symmetric, there is a unique
invariant coupling of four fermions to the $Sp(n)$ curvature.

In the context of Calabi-Yau compactification, there are
$h_{21}+1$ hypermultiplets, one for each pair of harmonic 3-forms. The
five-dimensional massless fermions $\c^I$ arise from the gravitino
zero-modes proportional to:
$$
\Psi^I_{0M}=d_{MNP}^I \G^{NP} (\e_+ +\e_-) \qquad{\rm with} \qquad
I=1,\dots,2h_{21}+2,
\eqn\zmdwt
$$
where $M,N=1,\dots,6$ and $d^I$ is a symplectic basis
of real harmonic three-forms, which obey:
$$
\int d^I \wedge d^J =\e^{IJ},
\eqn\dde
$$
with $\e^{IJ}$ the invariant antisymmetric tensor of $Sp(h_{21}+1)$.
\section{Membrane Zero-Modes, Vertex Operators and the Quantum Corrected
Geometry}
\no Membrane fermion zero-modes are generated by supersymmetries which are
unbroken
by the Calabi-Yau compactification, but broken by the presence of the
supermembrane. For the membrane instanton there are four such Nambu-Goldstone
zero-modes, given by
$$
\T_0(\s) =\ep_0(X(\s))
\eqn\zmd
$$
where
$$
P_+\ep_0=0.
\eqn\zmdt
$$

In general there may be additional zero-modes. This will indeed
be the case if the instanton is part of a continuos family, since
bosonic zero-modes must come with fermionic superpartners. In this paper
we consider only the minimal case of four zero-modes.

We will compute the contribution of membrane instantons to the low-energy
five-dimensional effective action. Since there are four fermion zero-modes, the
simplest quantity to compute is the correction to the couplings {\fffm} of four
space-time fermions. Corrections to other terms then follow from supersymmetry.

The low-energy effective field theory describing a membrane soliton coupled to
eleven-dimensional supergravity contains bilinear couplings of
the space-time fermions to membrane fermions. In particular there is a coupling
to the gravitino
$$
\int d^3 \s \sqrt{h}{\bar \Psi}_M (X(\s)) {\cal V}^M (\s),
\eqn\gurt
$$
where ${\cal V}^M$ is the membrane vertex operator\footnote{8}{The usual
string vertex operators can be derived in a similar manner by treating the
fundamental string as a soliton.}. This operator is uniquely fixed by
coordinate and $\k$-invariance as
$$
{\cal V}^M=\left(h^{\a\b} \pa_{\a} X^M \pa_{\b} X^N \G_N-
{i\o 2} \e^{\a\b\g} \pa_{\a} X^M \pa_{\b} X^N \pa_{\g} X^P \G_{NP} \right) \T,
\eqn\urt
$$
to leading order in the fermionic variable $\T$, at zero momentum,
and up to an overall normalization which can be deduced from formulae
in ref. [\smb] (we suppress
normalization constants in subsequent equations).

The vertex operator associated to the five-dimensional field $\chi^I$ is then
obtained from the zero-mode wave function {\zmdwt}:
$$
{\cal V}^I={\bar \Psi}^I_{0M}{\cal V}^M,
\eqn\xab
$$
and there is a corresponding coupling ${\bar \chi}^I {\cal V}_I$.
This coupling
leads to a non-vanishing instanton contribution to the correlator
$\langle \left( {\bar \c}_I \c_J\right)\left({\bar \c}_K
\c_L\right)\rangle$ (where indices are lowered with
$\e_{IJ}$), corresponding to an instanton-induced
correction to the effective action proportional to
$$
\D_{{\cal C}_3}{\cal R}_{IJKL}=\langle ({\bar \chi}_I \chi_J )
({\bar \chi} _K \chi_L)
\rangle_{\scriptstyle Instanton}.
\eqn\stcr
$$
The four $\T$ zero-modes can be absorbed by pulling down four powers of
${\bar \chi}^I {\cal V}_I$ from the action. The four left over
space-time fermions
are then contracted with the four fermions in eq. {\stcr}.
One therefore finds that
the space-time correlator {\stcr} is reduced to a correlator
in the three-dimensional field theory on the membrane
$$
\langle( {\bar \chi}_I \chi_J)(
{\bar  \chi} _K \chi_L)\rangle_{\scriptstyle Instanton}
=\langle({\bar {\cal V}}_I {\cal V}_J )(
{\bar {\cal V}}_K {\cal V}_L)
\rangle_{\scriptstyle Instanton},
\eqn\mbcr
$$
which has just the right number of $\T$'s to soak up the four fermion
zero-modes on the membrane. To evaluate this, we substitute the four zero-mode
wave functions
{\zmd} into the vertex operators:
$$
\int d^3 \s \sqrt{h} {\cal V}^I=\int d^3\s \sqrt{h}  \e^{\a\b\g} \pa_{\a} X^M
\pa_{\b} X^P \pa_{\g} X^Q d_{MRS}^I ({\bar \e}_++{\bar \e}_-) \G^{RS}
\G_{PQ}(\e_++\e_-).
\eqn\ver
$$
Using the identities of the appendix and {\jpl} one finds
$$
\int d^3 \s \sqrt{h} {\cal V}_I =\int\nolimits_{{\cal C}_3} d_I,
\eqn\eiv
$$
where ${\cal C}_3$ is the homology class of the instanton. Weighting by the
bosonic membrane action for the instanton one thereby obtains
$$
\Delta_{{\cal C}_3} {\cal R}_{IJKL} =N
 e^{-e^{-\cal K} |\int\nolimits_{{{\cal C}}_3}\O |-i
\int\nolimits_{{{\cal C}}_3}C}
\int\nolimits_{{{\cal C}_3}}d_I  \int\nolimits_
{{\cal C}_3}d_J \int\nolimits_{{\cal C}_3}d_K
\int\nolimits_{{\cal C}_3}d_L,
\eqn\ev
$$
The prefactor $N$ here and hereafter is inserted to remind the
reader that we have not computed the normalization factors, which also
includes a determinant here.
This formula indicates that membrane instantons give non-perturbative
quantum corrections to the modular geometry similar to the non-perturbative
classical $\a'$ corrections of string instantons
[\dsww]. To obtain the full corrections one must sum contributions
of the type {\ev} over all supersymmetric cycles, suitably modified to
account
for the possible effects of additional fermion or boson zero-modes
as well as multiple covers of the Calabi-Yau 3-cycle by the membrane.

\section{Relation to Four-Dimensional Extremal Black Holes}
\no The results of the previous sections can be used to partially
fill a gap in the discussion of ref. [\mbh]. In that work,
it was argued that the threebranes of the type IIB theory can
wrap around 3-cycles and appear as four-dimensional black holes; this is
in contrast to the wrapping modes of IIA twobranes which give instantons.
Four dimensional supersymmetry implied a bound on the black hole mass
of the form
$$
M\geq e^{{\cal K}_{\cal H}/2} |\int \O|.
\eqn\meo
$$
On the other hand, the threebrane in ten dimensions has a fixed
mass per unit three-volume. Thus one expects $M\propto V_3$, where
$V_3$ is the volume of the threebrane. The minimal threebrane volume
should be bounded by the period
of $\O$. This will indeed be the case if the conditions for a
supersymmetric IIB threebrane soliton are the same
as for a supersymmetric IIA
twobrane instanton. Verification of this will require a better
understanding of the threebrane dynamics, but we expect that solutions
of {\jpl} and {\opl} count extremal black holes in the IIB
theory as well as IIA instantons.

\chapter{Non-Perturbative Effects in Type IIA String Theory}
\no In this section non-perturbative effects in type IIA string theory
compactified to four dimensions on a Calabi-Yau manifold are derived.
These arise from RR instantons corresponding to the
twobrane solution of ref.
[\hs] wrapping around a 3-cycle of the Calabi-Yau space, and
correct the geometry of the hypermultiplet moduli space.

The low-energy effective action of the type IIA string theory is the non-chiral
$N=2$ and $D=10$ supergravity, which was originally derived
\REF\fen{I.~C.~G.~Campbell and P.~West, ``$N=2$ $D=10$ Non-Chiral Supergravity
and its Spontaneous Compactification'', \np {\bf B243} (1984) 112; F.~Giani
and M.~Pernici, ``$N=2$ Supergravity in Ten-Dimensions'', \pr {\bf D30}
(1984) 325; M.~Huq and M.~A.~Namazie, ``Kaluza-Klein Supergravity in
Ten Dimensions'', {\it Class. Quant. Grav.} {\bf 2} (1985) 293.}
[\fen]
by reduction of $D=11$ supergravity. The effective action of the
membrane solitons of ref. [\hs], is simply the supermembrane action {\sma}.
Instanton effects in the IIA theory can thus be derived by $S^1$
compactification of the eleven-dimensional analysis of the previous section.
The extra $S^1$ leads to a second type of membrane instanton, in which the
membrane
 wraps around the $S^1$ and a Calabi-Yau 2-cycle. In accord with the
observations that IIA strings are compactified $D=11$
membranes
\REF\dhis{M.~J.~Duff, P.~S.~Howe, T.~Inami and K.~S.~Stelle, ``
Superstrings in $D=10$ from Supermembranes in $D=11$'', \pl {\bf B191} (1987)
70.}
[\dhis,\town] and that eleven dimensional supergravity is strongly-coupled IIA
string theory [\wit], we shall see that in this case our formulae reduce to
standard worldsheet instanton corrections. Membranes which wrap Calabi-Yau
3-cycles lead to new non-perturbative string effects which behave as
$e^{-1/g_s}$.

\section{IIA Membrane Instantons}

\no Reduction to $D=10$ proceeds by periodically identifying
$$
X^{11}\simeq X^{11}+1 ,
\eqn\ide
$$
in eq. {\sma}. The $D=11$ spinor $\T$ is decomposed into two $D=10$
spinors $\T_{\pm}$ obeying $\G_{11}\T_{\pm}=\pm \T_{\pm}$. The
eleven-dimensional metric ${\widehat g}$ is related to the ten-dimensional
string metric $g$ and dilaton $\phi$ by [\fen,\wit]
$$
d{\widehat s_{11}}^2=e^{4\phi /3} \left(dx_{11}+A_m dx^m \right)^2
+e^{-2\phi /3}ds_{10}^2,
\eqn\liel
$$
where $A$ is a one-form.
The eleven-dimensional supergravity action then reduces to
$$
\eqalign{
{\cal S}_{10}=&
{1\o 2\pi^2  } \int d^{10} x \sqrt{g}
\left[ e^{-2\phi}\left(-R-4\left( \nabla \phi \right)^2 +{1\o 12}(dB)^2
\right) +{1\o 4}
(dA)^2 +{1\o 48} J^2 \right] \cr
&+{i \o 4\pi^2 }  \int  dC\wedge dC \wedge B,\cr}
\eqn\tenac
$$
where $B$ is the 2-form obtained
by the reduction $C=dx^{11}
\wedge B $ and $J=dC+A \wedge dB$ and our units are $\ell^2=2\pi \a'=1$.
For membranes which wrap Calabi-Yau 3-cycles, the instanton corrections
to hypermultiplet geometry are obtained simply substituting the appropriate
field redefinitions into eq. {\ev}. In particular
$$
{\wid V}_6={1\o 3!} \int {\wid J} \wedge {\wid J}  \wedge {\wid J} ,
\eqn\volu
$$
in eq. {\ev} is the volume of the Calabi-Yau space as measured by the
eleven-dimensional supergravity metric.
This is related to the string frame volume by
$$
{\widehat V}_6={1\o g_s^2 }V_6.
\eqn\volu
$$
Eq. {\ev} then becomes
$$
\Delta_{{\cal C}_3}
 {\cal R}_{IJKL} =N
e^{-{1 \o g_s} e^{-{\cal K}} |\int\nolimits_{{\cal C}_3}\O |-
i\int\nolimits_{{\cal C}_3}C}
 \int\nolimits_{{\cal C}_3}d_I  \int\nolimits_
{{\cal C}_3}d_J \int\nolimits_{{\cal C}_3}d_K  \int\nolimits_{{\cal C}_3}
d_L ,
\eqn\ronce
$$
where it is understood that the ${\cal K}$ in this expression
is constructed from the string K\"ahler form $J$ rather than ${\wid J}$.
The fact that non-perturbative corrections in string theory
should behave like $e^{-1/g_s}$ was predicted by Shenker [\shenk] from an
analysis
of the growth of the perturbation expansion. It is satisfying to see
this can be realized in an explicit calculation, as suggested in ref. [\wit].

\section{IIA String Instantons}

\no There are also membrane instantons which wrap the $S^1$ and a Calabi-Yau
2-cycle, i.e.
$$
X^{11}=\s_3,
\eqn\msr
$$
while $X^m$, $X^{\bar m}$ are functions of $\s_1$ and $\s_2$. These instantons
will correct the moduli space geometry of the vector multiplets. A slight
modification of the analysis of the previous section is required to find the
supersymmetric maps.
Using eq. {\msr} the membrane action {\sma} reduces to the
Green-Schwarz string action [\dhis]
$$
{\cal S}_2={1\o 2}
\int d^2 \s \sqrt{h}\left[h^{\a\b} \pa_{\a} X^M
\pa_{\b}X^N g_{MN}+i\e^{\a\b} B_{MN} \pa_{\a} X^M
\pa_{\b}X^N+{\rm (fermi)}\right].
\eqn\me
$$
Note that the factors of $e^{\phi}$ cancel when the action
is written in terms of the string metric.
The $\k$-transformations of $\T$ reduce to
$$
\d_{\k_{\pm}} \T_{\pm}=2 P_{\pm} \k_{\pm},
\eqn\kktpm
$$
where
$$
P_{\pm}={1\o 2} \left( 1\pm { i\o 2 } \e^{\a\b} \pa_{\a} X^M\pa_{\b} X^N
\G_{MN}\right).
\eqn\mee
$$
Repeating the analysis of the previous section, one finds that a
global supersymmetry transformation is equivalent to a $\k$-transformation
if and only if
$$
{\bar \pa} X^m=0 \qquad {\rm or} \qquad \pa X^{\bar m} =0,
\eqn\adr
$$
where ${\bar \pa}$ is the antiholomorphic worldsheet exterior derivative.
Eq. {\adr} is of course the usual holomorphy condition for worldsheet
instantons.

The Bogomol'nyi bound for the
induced area is now expressed in terms of the K\"ahler form as
$$
V_2\geq \int J,
\eqn\bvi
$$
and it is saturated by states satisfying {\adr}.

The vertex operators of the ten-dimensional theory can be
easily obtained from a double reduction of eq. {\urt}
$$
{\cal V}^M_{\pm}=\left(h^{\a\b}\mp i \e^{\a\b} \right)  \pa_{\a} X^M \pa_{\b}
X^P \G_P  \T_{\pm} \qquad {\rm with} \qquad M,P=1,\dots,10.
\eqn\sver
$$
It is straightforward to show that {\sver} is invariant
under $\k$-transformations.
The solutions of the six dimensional Dirac equation which lead to
vector multiplets will now be written
in terms of the harmonic $(1,1)$ forms
$$
u^I_M=b_{MN}^I \G^N \left( \e_++\e_-\right)
 \qquad {\rm with} \qquad I=1,\dots,
h_{11},
\eqn\uone
$$
and $M,N=1,\dots, 6$. One finds that the associated
zero momentum vertex operators are given by
$$
\int  d^2 \s \sqrt{h} {\cal V}_I=\int\nolimits_{{\cal C}_2} b_I.
\eqn\intv
$$
The result for the four-point correlation function of the
fermionic operators {\sver} is therefore
$$
\D_{{\cal C}_2} {\cal F}_{IJKL}=N
 e^{-\int
\nolimits_{{\cal C}_2} J-i\int\nolimits_{{\cal C}_2} B}
\int\nolimits_{{\cal C}_2} b_I
\int\nolimits_{{\cal C}_2} b_J \int\nolimits_{{\cal C}_2} b_K
\int\nolimits_{{\cal C}_2} b_L.
\eqn\scorr
$$
For vector multiplets, the four-fermi coupling is the derivative
of the ``Yukawas'' ${\cal F}_{IJK}$ with respect to the modulus
$z^I$, where
$$
J+iB=z^I b_I.
\eqn\jbz
$$
Integrating {\scorr} once with respect to the modulus, one
recovers the standard formula for worldsheet instanton
corrections to Yukawas. This provides an important check on our derivations
(for the case of simple covers and minimal fermion zero modes).

Our analysis also gives insight into the relation between $N=2$, $D=4$ and
$N=1$, $D=5$ supergravity. For hypermultiplets, there is no difference in
the four- and five-dimensional geometries. However for vector multiplets
there is a marked difference. In five dimensions the geometry is
completely determined by the {\it constants} $C_{IJK}$ [\gst],
which for the case of Calabi-Yau compactified $D=11$ supergravity are
just the cubic intersection form [\aswit]. In four dimensions, however,
the $C_{IJK}$ are functions of the moduli, which characterize a special
geometry
[\dlwp].
This is in perfect harmony with the fact that membrane instantons can not
correct $D=5$ vector multiplet geometry because there are no 3-cycles
associated to the vector multiplet geometry. However such 3-cycles appear
upon $S^1$ compactification from $D=5$ to $D=4$, and non-perturbative
corrections can and do arise. Thus we see again that
the low-energy supergravity theories ``know'' about non-perturbative
corrections!

\section{A Puzzle}
\no It was discovered in ref.
[\dhis] that, when eleven-dimensional supergravity is $S_1$-compactified
to IIA supergravity, a supermembrane with one leg wound
around the $S_1$ becomes the IIA string. This observation was used in the
previous subsection to reduce certain eleven-dimensional membrane
instantons to IIA string instantons.

However one may also consider a membrane for which one leg winds not once but
$n$ times around the $S_1$. This leads to ten-dimensional strings
with $n$ times the minimal string tension and axion charge. Such objects
cannot be incorporated in to IIA string theory.

The rules for inclusion of such configurations must be determined by
consistency.
One possible out is to regard the $n$-wound
membrane
as $n$ fundamental IIA strings on top of one another
rather than as a fundamental charge $n$ object. Superficially this is
similar to the situation encountered in ref. [\mbh], in which
charge $n$ black holes were treated as $n$ minimal charge black holes.
However subtleties arise in the present context. For a compactification
to $D=9$ on a second $S_1'$, there are Dabholkar-Harvey winding states
of the IIA string around $S_1'$. An $n$-wound string is {\it not}
the same as $n$ single-wound strings: there is a
bound state at threshold
\REF\dabhar{A.~Dabholkar and J.~A.~Harvey, ``Nonrenormalization
of the Superstring Tension'', \prl {\bf 63} (1989) 478.}
[\dabhar]. From the eleven-dimensional perspective these states are toroidal
membranes with winding $(1,m)$ around $(S_1,S_1')$. If a state
with winding number $(1,m)$ is allowed then by eleven-dimensional
Lorentz invariance $(n,1)$ should be allowed as
well. Conceivably $(n,1)$ and $(1,n)$ are identified by modular transformations
of the torus. Similar issues arise in counting multiple covers of
membrane/string instantons. We will not attempt to resolve the issue here.
We merely wish to emphasize that it is unresolved.

\chapter{Fivebrane Instantons}
\no In addition to strings and membranes, type II string theories contain
fivebrane solitons, which are described by exact conformal field theories
\REF\wsheet{C.~G.~Callan, J.~A.~Harvey and A.~Strominger, ``Worldsheet
Approach to Heterotic Instantons and Solitons'', \np {\bf B359} (1991) 611. }
\REF\wbran{C.~Callan, J.~A.~Harvey and A.~Strominger, ``Worldbrane
Actions for String Solitons'', \np {\bf B367} (1991) 60.}
[\hetssl,\wsheet,\wbran].
Euclidean fivebranes wrapping around a Calabi-Yau
space lead
to non-perturbative corrections to hypermultiplet geometry which in
principle might be computed analogously to string and membrane
instanton corrections. However, while the static gauge worldsheet field
content was derived in ref. [\wbran], the covariant action has not been
constructed.
Fortunately there is a much simpler way to proceed. Explicitly
the neutral four-dimensional field configuration is
$$
\eqalign{
H_{\m\n\l}&=-2{\e_{\m\n\l}}^\r\nabla_{\r} \phi, \cr
g_{\m\n}&=e^{2\phi} \d_{\m\n},\cr
e^{-2\phi}&=e^{-2\phi_0}+{1 \o 2\pi r^2}\cr}
\eqn\hge
$$
where $H=dB$.
The internal six-dimensional geometry is unaffected by the fivebrane
instanton, so standard four-dimensional instanton methods can be used to
compute corrections from {\hge}. The instanton action is \footnote{9}{The $a$
dependence of the action arises from a boundary
term which must be added to the action to fix the asymptotic value
of $a$. A very similar case is treated in ref.
\REF\sgas{S.~B.~Giddings and A.~Strominger, ``String Wormholes'',
\pl {\bf B230} (1989) 46.}[\sgas]. The strange-looking factor of
$1/\pi$ is the price we pay for using units in which the string
and membrane tensions are unity.}
$$
{\cal S}_6={1\o \pi g_s^2}+{ia\o \pi} ,
\eqn\acin
$$
where the axion field $a$ is related to $H$ by

$$
da={1\o 2} e^{-2\phi}*H.
\eqn\axion
$$

This instanton has four zero-modes obtained from
supersymmetry, for which the dilatino wave functions are
proportional to
$$
\c^D=\nsl \phi \eta_+.
\eqn\chzmo
$$
$\c^D$ is the linear combination of the ${\c^I}$'s corresponding
to the dilatino.
This leads to instanton corrections to the coupling of four dilatinos
$$
\D_{{\cal C}_6}{\cal R}_{DDDD}= e^{-1/\pi g_s^2-ia/\pi }+{\rm c.c.}.
\eqn\diin
$$
Similar corrections arise in $D=11$ supergravity and IIB string theory.
In the latter case, the formulae must be modified when the second RR dilaton
has non-zero expectation value. Furthermore, there is a second fivebrane
[\wbran] which couples to the RR Kalb-Ramond field.

\chapter{Mirror Symmetry and the IIB String}
\no At string tree level, mirror symmetry implies an exact equivalence between
the IIA string theory compactified on a Calabi-Yau space ${\cal X}$ (with
$N_{\cal V}=h_{11}$ vector multiplets and $N_{\cal H}=h_{21}+1$
hypermultiplets) and
IIB string theory (with $N_{\cal V} =h_{21}$ and $N_{\cal H} =h_{11}+1$) on the
mirror
${\widetilde {\cal X}}$ of ${\cal X}$. It is natural to suppose that these
theories are also equivalent at the quantum level. In particular
the instanton corrections to hypermultiplet geometry should be the same in
each case.

While the IIA theory contains RR membranes and fourbranes the IIB theory
contains RR strings, threebranes and fivebranes. There are generically no
interesting
fourbrane instantons which would have to wrap 5-cycles. Thus quantum mirror
symmetry will relate the sum {\ev} over 3-cycles in the
IIA theory on ${\cal X}$ to a sum over even-cycles in the IIB theory
on ${\widetilde {\cal X}}$.

The threebrane solution was discovered in ref. [\hs], and
its static gauge field
content was found in ref.
\REF\duffthree{M.~J.~Duff and J.~X.~Lu, ``The Self-Dual Type IIB
Superthreebrane'', \pl {\bf B273} (1991) 409.}
[\duffthree] to be an abelian $D=4$, $N=4$ vector multiplet.
Unfortunately the covariant action is as yet unknown so some guesswork will
be needed to deduce the IIB analogue of eq. {\ev}.
The bosonic part of the covariant action for $F=0$ is presumably
$$
\eqalign{
{\cal S}_4=\int & d^4 \s \sqrt{h}\Big( {1\o 2} h^{\a\b} \pa_{\a} X^M \pa_{\b}
X^N
e^{-\phi} g_{MN} -1 \cr
& +{i\o 4!} \e^{\a\b\g\d} D_{MNPQ} \pa_{\a} X^M
\pa_{\b} X^N \pa_{\g} X^P \pa_{\d} X^Q\Big), \cr}
\eqn\sfn
$$
where $D$ is a closed rank four antisymmetric potential. Note the factor of
$e^{-\phi}$ which arises because, as follows from formulae in
ref. [\hs]\footnote{10}{The threebrane geometry
 was computed in the Einstein frame
in ref. [\hs], the factor of $1/g_s$ is seen after rescaling to the
(fundamental) string frame.}, the action per unit four-volume of a
charge one threebrane is
inversely proportional to the string coupling.

The real part of the action is proportional to the four-volume
of the threebrane.
It is not hard to show that the volume satisfies the bound
$$
V(\S_4)\geq {1\o 2} \int J \wedge J.
\eqn\vfbd
$$
This bound is saturated if the map $X(\s)$ is holomorphic
$$
{\bar \pa} X^m(\s,{\bar \s})=0,
\eqn\hmp
$$
where we have introduced complex coordinates on the threebrane. We presume that
the configurations which satisfy {\hmp} preserve half of
the supersymmetries. In addition there are abelian gauge fields on the
worldbrane. The structure of the ten-dimensional supersymmetry
transformations suggest that self-dual gauge configurations will break
some, but not all, of the supersymmetries preserved by {\hmp}. Since
more supersymmetries are broken by such configurations, there will be
more than four fermion zero-modes. Such configurations
might still contribute to the four-fermi couplings but will not be further
considered.

If the threebrane instanton breaks half the supersymmetries, there will be
four Goldstino zero-modes and corrections to four-fermi couplings.
In analogy with the string and the membrane case, the natural expression
for these corrections is
$$
\D_{{\cal C}_4} {\cal R}_{IJKL}= N e^{-{1\o 2 g_s}
\int\nolimits_{{\cal C}_4} J
\wedge J -i \int\nolimits_{{\cal C}_4} D}
\int\nolimits_{{\cal C}_4}f_I
\int\nolimits_{{\cal C}_4}f_J \int\nolimits_{{\cal C}_4}f_K
\int\nolimits_{{\cal C}_4}f_L.
\eqn\tbhm
$$
${\cal C}_4$ is a holomorphic map and $f_I$ is the four form
Hodge dual to the I'th (1,1) class. Of course with
no restriction on topology {\tbhm} will be
hard to evaluate. It is possible that, as in the string case, an
analysis of zero-modes will lead to a restriction on the topology.

This is not the end of the story. IIB string theory contains a
second soliton
string
\REF\hast{J.~A.~Harvey and A.~Strominger,
``The Heterotic String is a Soliton'', \np {\bf B} to appear,
preprint EFI-95-16, hep-th/9504047.}
\REF\hullss{C.~M.~Hull, ``String-String Duality in Ten-Dimensions'',
preprint QMW-TH-95-25, hep-th/9506194.}
[\hast,\hullss]
which is related to the fundamental IIB string by an $SL(2,\IZ)$
duality transformation. Worldsheet instantons of the soliton string
will give corrections to hypermultiplet geometry similar to those of
fundamental strings, except with two important differences:
\item{a)} the string tension and therefore the instanton action differs by a
factor of $1/g_s$.
\item{b)} it couples to the second Kalb-Ramond field, which we shall denote by
${\widetilde B}$, arising in the RR sector.

\no The corrections are:
$$
\D_{{\cal C}_2} {\cal R}_{IJKL}=N
e^{-{1\o g_s} \int
\nolimits_{{\cal C}_2} J-i\int\nolimits_{{\cal C}_2} {\widetilde  B}}
\int\nolimits_{{\cal C}_2} b_I
\int\nolimits_{{\cal C}_2} b_J \int\nolimits_{{\cal C}_2} b_K
\int\nolimits_{{\cal C}_2} b_L .
\eqn\rtwo
$$
$SL(2,\IZ)$ invariance implies that a genus $g$ instanton is weighted by
$(g_s)^{-2g}$, but it also implies that $g\geq 0$ instantons
will not correct hypermultiplet geometry.

Finally there is a second fivebrane, which couples to the dual of the
RR Kalb-Ramond field strength, and has an action which,
in the string frame, is greater by a factor of $g_s$ than the NS-NS fivebrane.
Therefore it corrects the four-dilatino coupling by:

$$
\D_{{\cal C}_6}{\cal R}_{DDDD}=N e^{-1/\pi g_s}.
\eqn\rsix
$$
There is presumably also some exponential dependence on other
members of the dilaton hypermultiplet (this could also arise
in eqs. {\tbhm} and {\rtwo}) which we have not analyzed.

Quantum mirror symmetry then relates the IIB sums of $e^{-1/g_s}$
corrections {\tbhm},
{\rtwo} and {\rsix}, to the IIA sum {\ev}. It may be useful at this point to
summarize the main steps required
to obtain a precise mathematical formula. On the IIA side,
we must understand when (if ever) there are more than four zero-modes,
and determine the contributions of such 3-cycles. On the IIB side, we
must verify {\tbhm}, presumably by first constructing the covariant
threebrane action. The zero-mode and multiple cover
problems must then be analyzed. There
is also the possibility that contributions to {\tbhm} are weighted
by some power of $g_s^{\chi}$, where $\chi$ is the Euler character
of the surface. Phases associated with the signature of the surface may
also appear. There may be ambiguities in relating the coordinates
on the two moduli spaces. Finally one hopes to evaluate this sums
and check the predictions in special cases or limits. Clearly much
remains to be done. We hope to report on these issues elsewhere.

\chapter{Conifold Singularities in the Hypermultiplet Moduli Space}
\no The moduli space of a Calabi-Yau space generically contains
conifolds at which the moduli space metric becomes singular. If the
moduli lie in vector multiplets, this singularity is resolved by
charged
BPS states which become massless at the conifold [\mbh,\tbh].
Such a resolution
can not occur for Calabi-Yau moduli which lie in hypermultiplets
because there are no associated charges,
BPS bounds or massless states. However, the consistency of string
theory
demands some resolution of the hypermultiplet singularities.
Since hypermultiplet geometry is not protected by a
non-renormalization
theorem, it is natural to suppose that the singularity is eliminated
by quantum corrections.

Some insight into the problem can be gained by considering a field
theory
analog. In particular $D=4$, $N=2$ $SU(2)$ Yang-Mills theory contains a
conifold singularity in the vector multiplet moduli space which is
resolved by light monopoles. After compactification to $D=3$ on a
circle, the monopoles become infinitely massive (due to long-range
fields) and leave the Hilbert space. Therefore they can no longer
resolve the singularity. This singularity is actually in a
hypermultiplet geometry, since in $D=3$ vector multiplets dualize
to hypermultiplets.
However, the $D=3$ theory contains instantons corresponding to $D=4$
monopoles orbiting the compactification circle. The instanton sum
contains a logarithmic divergence which seems to cancel the logarithm
of the four dimensional theory
\REF\witap{E.~Witten, private communication and to appear.}
\REF\shenker{S.~Shenker, ``Another Length Scale in String Theory?'',
preprint RU-95-53, hepth/9509132.}
[\witap,\shenker], although the details have not all been
worked out. Thus in the field theory example it appears that quantum
corrections resolve the singularity.

This analogy is in fact quite close due to the following observation
\REF\pol{J.~Dai, R.~G.~Leigh and J.~Polchinski, ``New Connections Between
String Theories'', \mpl {\bf A4} (1989) 2073.}
\REF\dhl{M.~Dine, P.~Huet and N.~Seiberg, ``Large and Small Radius in
String Theory'', \np {\bf B322} (1989) 301.}
[\pol,\dhl].
Compactification of the IIA string theory on a product of a
Calabi-Yau space ${\cal X}$ with a circle of radius $R$ is equivalent to
compactification on the mirror ${\widetilde{\cal X}}$ of ${\cal X}$ and a
circle of radius $1/R$.
In $D=4$, exchanging ${\cal X}$ with ${\widetilde{\cal X}}$ interchanges
vector multiplets and hypermultiplets.
Since hypermultiplet bosons are all scalars, compactification from
$D=4$ to $D=3$ does not change the hypermultiplet geometry.
The hypermultiplet geometry for the IIA theory on ${\cal X}$ can
therefore be found by compactifying the IIA theory on
${\widetilde{\cal X}}$ on a circle of radius $R$ and then taking $R$
to zero. Hence exactly the same mechanism which eliminates
hypermultiplet singularities in the $D=3$ field theory example
should eliminate them in $D=4$ type II string compactifications.
This remains to be verified or understood in detail.

\chapter{Fundamental $p$-Branes?}
\no A striking feature of our calculation is the democratic
fashion in which all
the $p$-branes enter. At the non-perturbative level,
the role played by strings does not seem special. Indeed, a glance
at the brane-scan
\REF\bsc{M.~J.~Duff and J.~X.~Lu,
``Type II $p$-Branes, the Brane-Scan Revisited'',
\np {\bf B390} (1993) 276, hep-th/9207060.}
[\tls,\bsc] reveals that strings occupy a relatively
undistinguished position, compared with the maximally symmetric
positions occupied by the membrane and fivebrane.
Recent work [\tbh] in which strings smoothly
transform into twobranes or threebranes further threatens the privileged
role of strings in string theory.

One might conclude from this that type II ``string theory '' is really
a theory of ``fundamental'' membranes and/or fivebranes
\REF\demo{P.~K.~Townsend, ``p-Brane Democracy'', preprint DAMTP-R-95-34,
hep-th/9507048.}
[\town,\smb,\demo].
However this is not our view. Rather we feel that string theory
can not be fundamentally defined as a theory of extended objects of any
kind -- including strings. The various different extended objects have
calculational utility in different regions of the moduli space. For
example we have seen that membranes are useful for understanding
aspects of strongly coupled, eleven-dimensional IIA theory [\wit] while
strings are not.  What is special about strings is that string diagrams
can be ordered by genus. This leads to a systematic perturbation
expansion of e.g. graviton-graviton scattering in certain regions of the
moduli space. It is doubtful that such calculation will be
possible in the near future
using membranes or fivebranes. But this does not mean that strings
are more ``fundamental'', they are just more useful.

What then is ``string theory''? We do not know, but in any case
it does not seem to be fundamentally defined as a theory of strings.

\vskip 2cm
\no {\bf Acknowledgements}

\no We are grateful to M.~J.~Duff, B.~Greene, S.~Ferrara, J.~Harvey,
M.~Kontsevich, G.~Moore, D.~Morrison,
J.~Polchinski, S.~Shenker, C.~Vafa,
E.~Witten and S.-T.~Yau for useful discussions.
We would like to thank the Aspen Center for Physics, the International
Center for Theoretical Physics of Trieste and Rutgers University
for hospitality where parts of this work have been carried out.
This work was supported in part by
DOE grant DOE-91ER40618 and NSF grant PHY89-04035.

\endpage
\no {\bf Appendix}

\no Our notation and conventions are as follows:
\item{\tria} $\G^N$ with $N=1,\dots,11$ are the euclidean space-time gamma
matrices and
$$
\G_{\a}=\pa_{\a}X^M\G_M\qquad {\rm with}\qquad \a=1,2,3
$$
are their projections onto the worldbrane. Using eq. {\ei} they satisfy
$$
\{\G_{\a},\G_{\b}\}=2h_{\a\b}
$$
\item{\tria}We use the notation
$$
\G^{M_1\dots M_n}=\G^{[M_1}\G^{M_2} \dots \G^{M_n]},
$$
where the square bracket implies a sum over
$n!$ terms with an $1/n!$ prefactor. \item{\tria}Six-dimensional gamma matrices
are chosen to satisfy $(\g_m)^*=
-\g_{\bar m}=\g_m^T $.
\item{\tria}$\e_{\a\b\g}$ transforms as a tensor under membrane
diffeomorphisms.
\item{\tria}For a spinor $\psi$ we define ${\bar \psi}=
\left(\psi^*\right)^T$ in euclidean space.
\item{\tria}$n$-forms are defined with a factor $1/n!$, so for example
$J={1\o 2} J_{MN}dX^M\wedge dX^N$.
\item{\tria}The following identity is useful to check the
$\k$-symmetry of our membrane vertex operator
$$
\G_{\a} P_{\pm}=\pm {i\o 2} {\e_{\a}}^{\b\g} \G_{\b\g} P_{\pm}.
$$
\item{\tria}To prove eq. {\eiv} the following identities are useful
$$
\eqalign{
{\bar \eta_+}\g^{r{\bar s}} \g_{p{\bar q}} \eta_+&= {\wid J}^{r{\bar s}}
{\wid J}_{p{\bar q}}, \cr
{\bar \eta_+} \g^{rs} \g_{pq} \eta_+&=8{{\wid J_p}}^{[r} {{\wid J_q}}^{s]}.
\cr}
$$
where $\eta_+$ and $\eta_-$ are two covariantly constant
six-dimensional spinors with
opposite chirality and unit norm.

\endpage
\refout
\end